\documentclass[%
 reprint,
 amsmath,amssymb,
 aps,
]{revtex4-2}

\usepackage{graphicx}
\usepackage{dcolumn}
\usepackage{bm}

\usepackage{color}

\begin{document}


\title{Dislocation avalanches from strain-controlled loading:\\
A discrete dislocation dynamics study}

\author{David Kurunczi-Papp}
\email{david.kurunczi-papp@tuni.fi}
\author{Lasse Laurson}%
\affiliation{%
 Computational Physics Laboratory, Tampere University, P.O. Box 692, FI-33014 Tampere, Finland
}%

\date{\today}

\begin{abstract}
We study strain-controlled plastic deformation of crystalline solids via two-dimensional discrete dislocation dynamics simulations. To this end, we characterize the average stress-strain curves as well as the statistical properties of strain bursts and the related stress drops as a function of the imposed strain rate and the stiffness of the specimen-machine system. The dislocation system exhibits strain rate sensitivity such that a larger imposed strain rate results in a higher average stress at a given strain. In the limit of small strain rate and driving spring stiffness, the sizes and durations of the dislocation avalanches are power-law distributed up to a cutoff scale, and exhibit temporally asymmetric average shapes. We discuss the dependence of the results on the driving parameters, and compare our results to those from previous simulations where quasistatic stress-controlled loading was used.
\end{abstract}

\maketitle


\section{\label{sec:1} Introduction}

Plastic deformation of crystalline solids is governed by stress-driven collective dynamics of dislocations~\cite{papanikolaou2017avalanches,alava2014crackling}, often exhibiting "complexity" manifesting itself as fluctuations of the deformation process, i.e., dislocation avalanches spanning a wide range of sizes~\cite{csikor2007dislocation,richeton2005dislocation}. For small, micrometer-sized crystals, these avalanches are visible directly as fluctuations and irregularities in the stress-strain curves, with quantities like strain increments/bursts and/or stress drops exhibiting broad, power-law-like size distributions~\cite{Dimiduk1188}. When deforming macroscopic samples, one often observes broadly distributed acoustic emission energies and amplitudes~\cite{PhysRevLett.114.105504,weiss2003three}, even if the stress-strain curves tend to be smooth. Besides the acoustic emission measurements, high-resolution extensometry experiments on Cu single crystals have been performed to analyse the statistics of the local strain rate and strain bursts~\cite{PhysRevB.79.014108}. In addition, dislocation plasticity is also dependent on crystal structure and orientation~\cite{PhysRevMaterials.2.120601}, and is characterized by size effects — typically in agreement with the paradigm “smaller is stronger”~\cite{uchic2004sample,dimiduk2005size,jennings2011emergence} — as well as rate effects, where the flow stress exhibits a dependence on the imposed strain rate (often referred to as "strain rate sensitivity") in strain-controlled simulations and experiments~\cite{fan2021strain,schwaiger2003some,jennings2011emergence,AGNIHOTRI201537}.

In order to grasp the origin and nature of the experimentally observed broadly distributed critical-like dislocation avalanches and related phenomena, a wide range of numerical work has been performed. Notable examples are given by discrete dislocation dynamics (DDD) simulations, considering either simplified two-dimensional (2D) geometries with point-like dislocations~\cite{ispanovity2014avalanches,ovaska2015quenched}, or more realistic 3D systems describing dislocations as flexible lines~\cite{lehtinen2016glassy,salmenjoki2020plastic}. These simulations have contributed to a theoretical picture where the key concepts include dislocation jamming~\cite{miguel2002dislocation,ispanovity2014avalanches,lehtinen2016glassy} for systems with negligible quenched disorder interacting with the dislocations, which are often found to exhibit slow, glassy dynamics, as well as (de)pinning transitions of dislocation assemblies in systems where strong enough static obstacles such as precipitates interfere with dislocation motion~\cite{ovaska2015quenched,salmenjoki2020plastic}. Another numerical model successfully reproducing the properties of dislocation systems is the integer-valued automaton representing the crystal as an array of Frenkel-Kontorova chains~\cite{https://doi.org/10.1002/pssb.2221790212,PhysRevLett.106.175503}. This computationally efficient 2D model, accounting for both short- and long range elastic interactions, including nucleation and immobilization, is capable of reproducing critical exponents of compression tests on Mo submicron pillars~\cite{SALMAN2012219,PhysRevE.102.023006}.

Many previous DDD simulations in 2D and 3D with the aim to model strain bursts in crystal plasticity have focused on stress-controlled loading where the applied stress $\sigma$ is increased quasistatically~\cite{ispanovity2014avalanches,ovaska2015quenched,lehtinen2016glassy,salmenjoki2020plastic}. This protocol leads to a staircase-like stress-strain curve $\sigma(\epsilon)$, consisting of horizontal segments (the strain bursts) where strain is accumulated at constant stress, separated by parts where the stress increases, producing quasireversible deformation~\cite{szabo2015plastic}. This quasistatic stress ramp loading has been demonstrated to result in power-law stress-resolved strain burst distributions $P(\Delta \epsilon;\sigma)$ with a stress-dependent upper cutoff $\Delta \epsilon_0 (\sigma)$, 
\begin{equation}\label{eq:powerlaw}
P(\Delta \epsilon;\sigma) \propto (\Delta \epsilon)^{-\tau_\epsilon} 
f\left[\frac{\Delta \epsilon}{\Delta \epsilon_0(\sigma)}\right],
\end{equation}
where $f$ is a scaling function, typically an exponential, $f(x)=\exp(-x)$. The exponent $\tau_\epsilon$ has been found to be close to $\tau_\epsilon \approx 1$ for 2D pure dislocation systems~\cite{ispanovity2014avalanches}. Moreover, the cutoff scale $\Delta \epsilon_0 (\sigma)$ is found to exhibit an exponential increase with $\sigma$, and a power law dependence on the number of dislocations $N$~\cite{ispanovity2014avalanches}. One may also consider the stress-integrated distribution $P_\mathrm{INT}(\Delta \epsilon) = \int P(\Delta \epsilon;\sigma) \mathrm{d}\sigma \sim (\Delta \epsilon)^{-\tau_{\epsilon,\mathrm{INT}}}$, which has been found to be characterized by a larger exponent value of $\tau_{\epsilon,\mathrm{INT}} \approx 1.3$ in stress-controlled 2D DDD simulations~\cite{ispanovity2014avalanches}. Similar results have been obtained also in stress-controlled 3D DDD simulations of pure Al single crystals~\cite{lehtinen2016glassy}. 

Here, we perform simulations of a 2D DDD model using strain-controlled loading, in order to characterize the dislocation avalanche statistics and related rate effects, and contrast the results with those from previous studies of the same and related models using quasistatic stress-controlled loading. When using strain-controlled loading, the stress-strain curves exhibit a sawtooth-like shape, consisting of a series of stress drops, coinciding with the strain bursts, separated by segments where $\sigma$ increases. We note that depending on the employed loading mode (stress vs strain-controlled~\cite{Kubin}), previous studies have observed different avalanche size scaling behaviors both theoretically and experimentally~\cite{PhysRevLett.101.230601, PhysRevB.80.180101}.

First, we demonstrate the system size dependence of rate-dependent plastic deformation. Our findings show that small systems are characterized by large fluctuations, the magnitude of which is reduced as the system size is increased~\cite{uchic2004sample}. Afterwards, having chosen a sufficiently large system, by considering the average stress-strain curves, we find a clear rate effect in the average flow stress at a given strain, i.e., the 2D DDD model exhibits non-zero strain rate sensitivity~\cite{schwaiger2003some,jennings2011emergence,AGNIHOTRI201537}. Interestingly, our results suggest that dislocation avalanches in the simple 2D DDD model appear to exhibit non-universal avalanche dynamics in that the statistical properties of the avalanches seem to depend on the parameters of the strain-controlled driving, i.e., the imposed strain rate and the stiffness of the specimen-machine system. This also implies that we find different avalanche exponents for the two loading modes, i.e., stress-controlled and strain-controlled loading. In the limit of small imposed strain rate and stiffness of the specimen-machine system, we find consistently larger exponent values for both strain burst and stress drop distributions than found previously for the strain bursts in quasistatic stress-controlled loading~\cite{ispanovity2014avalanches}. We discuss how different definitions of the avalanche threshold arising in the context of the two loading modes might give rise to these differences in observed avalanche scaling. Furthermore, we characterize the dependence of the average avalanche shape, including its temporal asymmetry, on the driving parameters. We find a clear temporal asymmetry exhibiting dependence on both the imposed strain rate and the stiffness of the specimen-machine system.

The paper is organized as follows: In Section~\ref{sec:2}, we present the DDD model used in our study, and show the results from numerical simulations of the model in Section~\ref{sec:3}, starting with the analysis of system size effects in~\ref{subsec:3.0}, and the characterization of the average stress-strain curves in~\ref{subsec:3.1}, followed by statistical analysis of stress drops and strain increments as well as event durations in~\ref{subsec:3.2}, and the average avalanche shapes in \ref{subsec:3.3}. The paper is finished with conclusions and discussion in Section~\ref{sec:4}.

\section{\label{sec:2} DDD Simulations}

The DDD model we consider is a simple and computationally efficient 2D model. The parallel, straight edge dislocations with equal number of positive and negative Burgers vector of magnitude $b$ are represented in the 2D Cartesian coordinate system as point-like objects moving along parallel lines in the $x$-direction. The dynamics of the dislocations is simulated in a square box of linear size $L$ with periodic boundary conditions, initially containing $N_0$ dislocations. The long-range shear stress field of a single dislocation is given by $\sigma_\mathrm{d}(\mathbf{r})=Db\frac{x(x^2-y^2)}{(x^2+y^2)^2}$ with the appropriate elastic constant $D$ as the prefactor~\cite{Dislocations}. Assuming overdamped dynamics where the dislocation velocity is proportional to the total Peach-Koehler force acting on it, the equation of motion of the $i$th dislocation along the $x$-direction becomes
\begin{equation}
\frac{\dot{x}_i}{Mb} = s_ib\left[\sum_{j\neq i}s_j\sigma_\mathrm{d}(\mathbf{r}_i-\mathbf{r}_j)+\sigma\right],
\end{equation}
where $M$ is the dislocation mobility, $s_i\in\{-1,1\}$ is the sign of the $i$th Burgers vector, and $\sigma$ the external shear stress. The strain-controlled loading is realized by adjusting the time-dependent external stress $\sigma(t)$ at each time step according to
\begin{equation}\label{eq:drive}
\sigma(t)=k\left[\dot{\epsilon}_at-\epsilon(t)\right],
\end{equation}
where $k$ is the driving spring stiffness ("stiffness of the specimen-machine system") and $\dot{\epsilon}_a$ the imposed strain rate. Notice that the driving protocol of Eq.~(\ref{eq:drive}) pushes the time-average of $\dot{\epsilon}$ towards the imposed strain rate $\dot{\epsilon}_a$, which also acts as a threshold level to define dislocation avalanches or strain bursts as events during which $\dot{\epsilon}$ exceeds $\dot{\epsilon}_a$. It should also be noted that the strain control in our simulations is not perfect, but an intermediate situation between "soft" (stress-controlled) and "hard" (strain controlled with an infinitely stiff machine) driving. As we consider different values of the specimen-machine stiffness, we are simulating systems with different levels of strain control, such that the limiting case of a vanishing stiffness would approach purely stress-controlled loading. However, the driving protocol used in our study will ensure that in the long-time limit the time-averaged strain rate will converge to the imposed strain rate, and in this sense strain is controlled in our simulations. We also compare the results from our strain-controlled simulations with ones using stress-controlled loading, where a small strain rate threshold is defined, and $\sigma$ is kept constant whenever the instantaneous $\dot{\epsilon}$ exceeds the threshold (i.e., during the strain bursts), and increased at a slow rate in between the bursts (when $\dot{\epsilon}$ is below the threshold).

\begin{figure}[t!]
	\centering
	\resizebox{0.95\columnwidth}{!}{\includegraphics{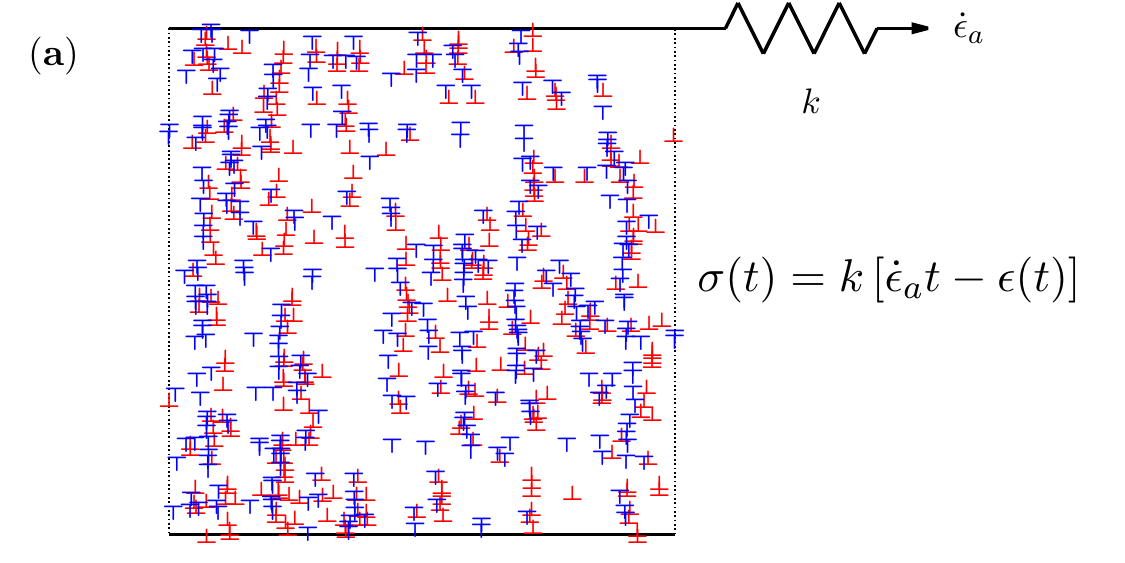}}
	\resizebox{0.95\columnwidth}{!}{\includegraphics{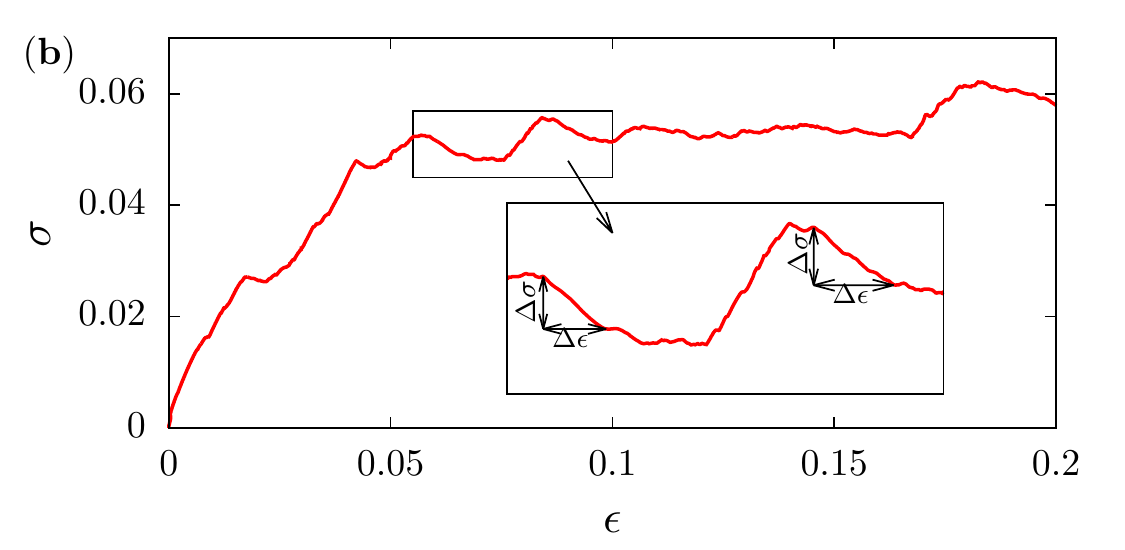}}
	\resizebox{0.95\columnwidth}{!}{\includegraphics{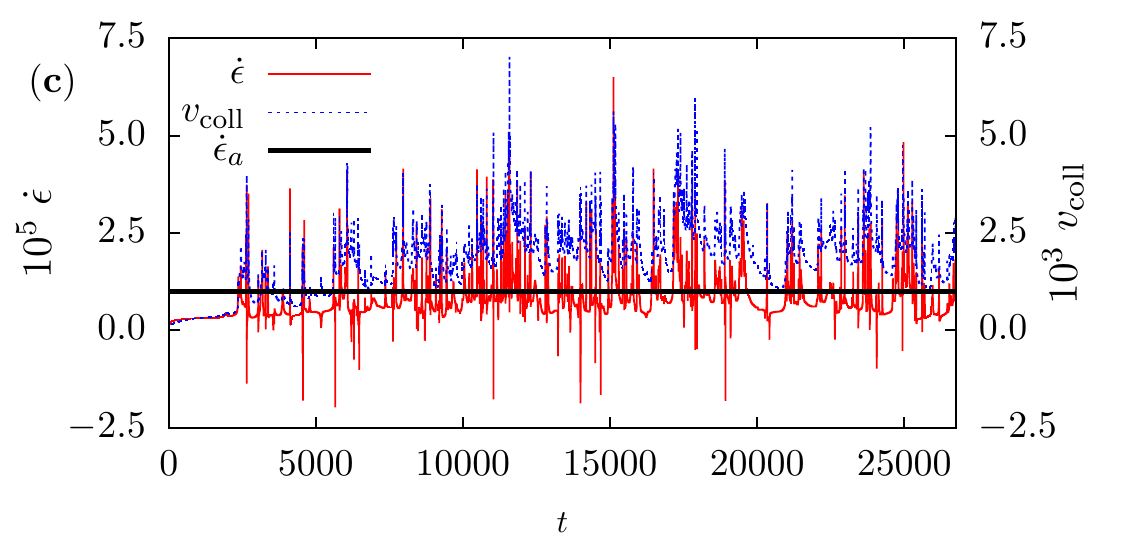}}
	\caption{(a) A schematic figure illustrating the simulation setup, with red and blue symbols corresponding to positive and negative Burgers vectors of the edge dislocations, respectively, under strain rate driven shear; periodic boundary conditions apply; (b) the resulting stress-strain curve for the parameters $k=1$ and $\dot{\epsilon}_a=10^{-5}$, indicating also the definitions of the monotonic stress drops $\Delta \sigma$ and the corresponding strain increments $\Delta \epsilon$, and (c) strain rate $\dot{\epsilon}$, collective velocity $v_{\mathrm{coll}}$ signals as a function of time and the imposed strain rate threshold $\dot{\epsilon}_a$.}
	\label{fig:1}
\end{figure}

In what follows, lengths, times and stresses are measured in  units of $b$, $(MDb)^{-1}$ and $D$, respectively. Fig.~\ref{fig:1}(a) illustrates the simulation box, including the dislocations with positive and negative Burgers vectors, respectively, and the strain-controlled loading with its governing equation. A resulting stress-strain curve is shown in Fig.~\ref{fig:1}(b), where the inset zooms on typical strain bursts and indicates the definitions of a monotonic stress drop $\Delta\sigma$ and its corresponding strain increment or strain burst $\Delta\epsilon$. The corresponding strain rate $\dot{\epsilon}=(b/L^2)\sum_i s_i \dot{x}_i$ and collective velocity $v_{\mathrm{coll}}=\frac{1}{N}\sum_i |\dot{x}_i|$ signals as well as the imposed strain rate $\dot{\epsilon}_a$ as a function of time are shown in Fig.~\ref{fig:1}(c). A notable difference between these two signals is that the strain rate can become negative, while the collective velocity is a strictly positive quantity. The strain rate signal $\dot{\epsilon}$ together with Eq.~\eqref{eq:drive} explains the occurrence of stress drops: $\sigma$ is a monotonically decreasing function of time whenever $\dot{\epsilon}>\dot{\epsilon}_a$.

To eliminate eventual singularities caused by the interaction of two dislocations very close to each other, annihilation of two dislocations with opposite Burgers vectors is simulated by removing both from the system when their distance becomes less than $b$~\cite{BAKO200622,PhysRevE.74.066106,PhysRevLett.105.015501,AGNIHOTRI201537}. Other features commonly implemented into 2D DDD simulations, such as immobile impurities accountable for dislocation pinning~\cite{Papanikolaou2012,PhysRevE.93.013309,PAPANIKOLAOU201717}, and dislocation climb or cross-slip allowing the dislocations to move in the $y$-direction \cite{BAKO200622,Papanikolaou2012} are not included into this simplified model. The lack of nucleation in our 2D DDD model is a non-negligible limitation, however real dislocation multiplication mechanisms such as Frack-Read sources~\cite{PhysRevE.74.066106,PAPANIKOLAOU201717} are fundamentally a property of curved, flexible dislocation lines, and hence they cannot be properly modelled in our 2D system, where the dislocations are understood to be straight lines with the line direction perpendicular to the simulation plane. These somewhat arbitrary choices of the simulation mechanisms are limitations of the 2D DDD models in general, and to properly describe such processes one would need to consider 3D DDD simulations. Here, considering the same 2D DDD model as in previous works where stress-controlled loading was applied~\cite{ispanovity2014avalanches, ovaska2015quenched} is justified in order to be able to compare our results with those studies.

In this study we consider initially random configurations of $N_0$ dislocations in a simulation box of linear size $L$ with the initial dislocation density fixed to $\rho_0\equiv N_0/L^2 = 0.04$. These random dislocation configurations are first let to relax at $\sigma=0$ for $t=15000$ to reach a metastable dislocation configuration containing various dislocation structures such as dislocation dipoles and walls. During this initial relaxation, approximately half of the dislocations get annihilated. After this initial relaxation stage, the loading protocol of Eq.~(\ref{eq:drive}) is switched on, and the resulting stress-strain curve is measured for each sample by storing the values of $\sigma$ and $\epsilon = (b/L^2)\sum_i s_i x_i$ during the simulations.

\section{\label{sec:3} Results}

\subsection{\label{subsec:3.0} System size effects}

\begin{figure}[t!]
	\centering
	\resizebox{0.95\columnwidth}{!}{\includegraphics{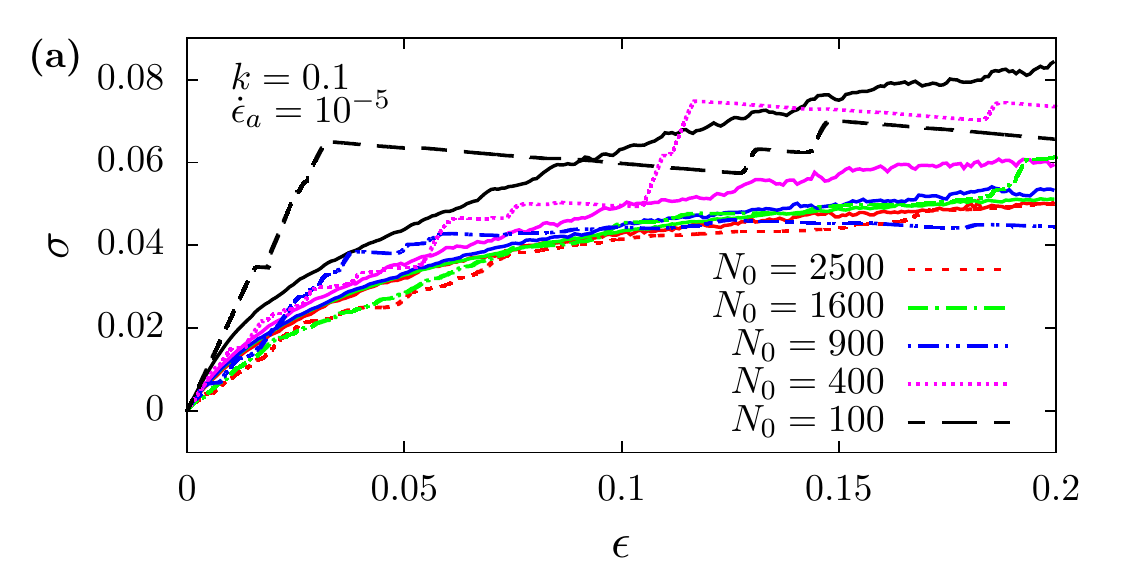}}
	\resizebox{0.95\columnwidth}{!}{\includegraphics{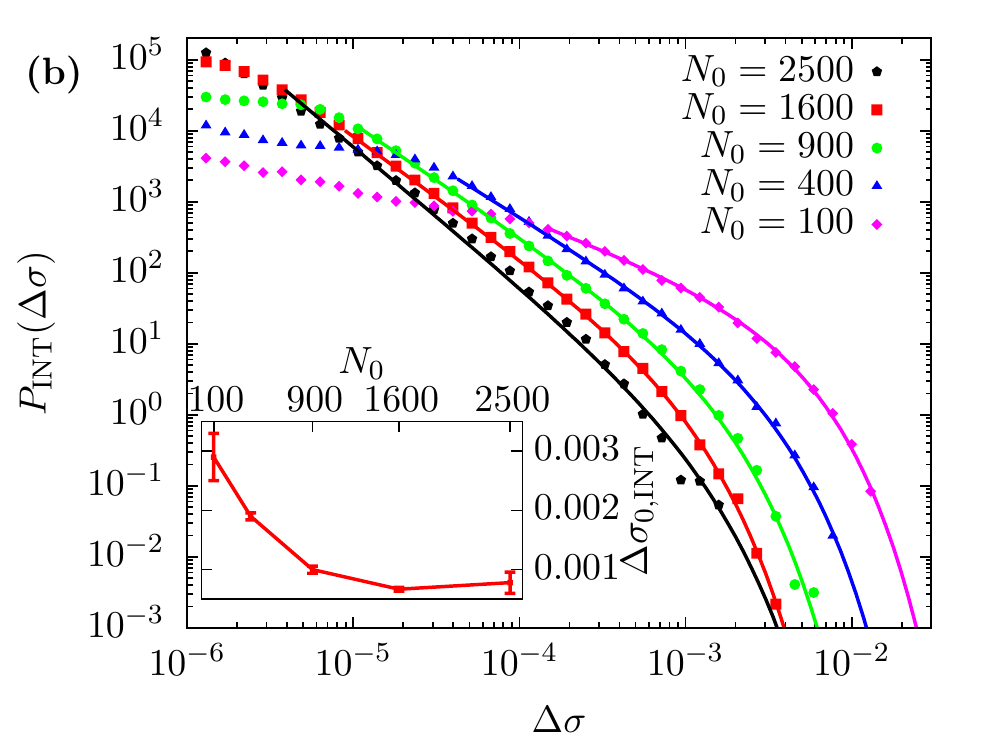}}
	\resizebox{0.95\columnwidth}{!}{\includegraphics{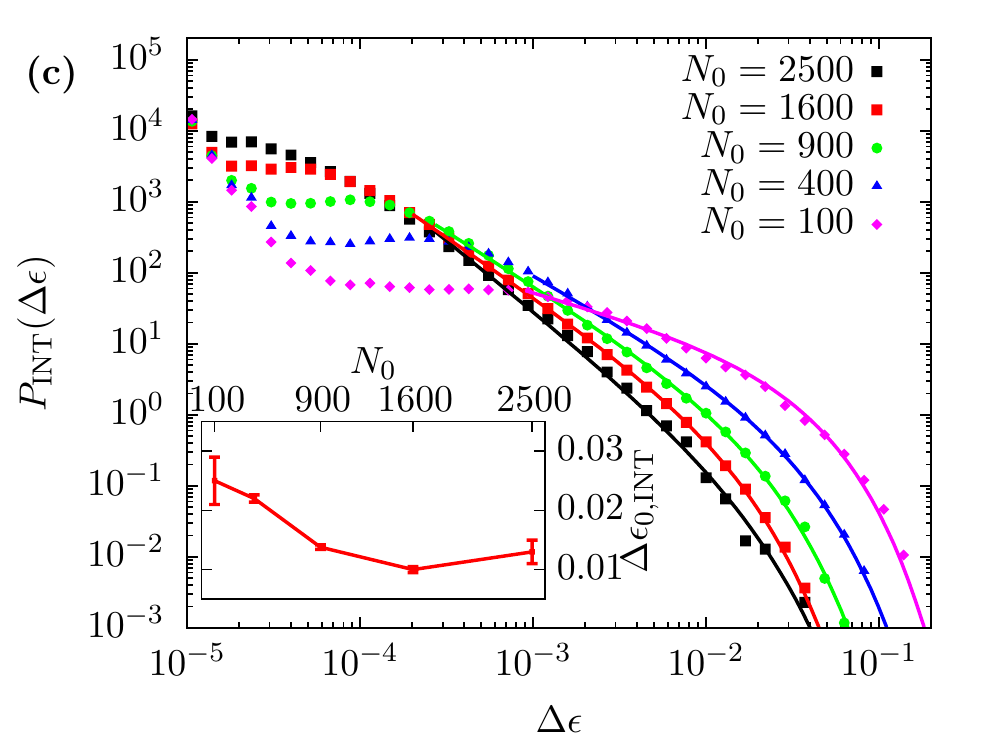}}
	\caption{(a) Average stress-strain curves $\langle \sigma(\epsilon) \rangle$ (with the average taken over different realizations of the initial configuration, solid lines) and examples of individual, single-sample stress-strain curves (patterned lines) for varying number of initial dislocations $N_0$ (the corresponding average curve increases with decreasing $N_0$), an imposed strain rate $\dot{\epsilon}_a=10^{-5}$ and driving spring stiffness $k=0.1$. Integrated distribution of the (b) stress drop magnitudes $P_{\mathrm{INT}}(\Delta \sigma)$ and (c) strain increments $P_{\mathrm{INT}}(\Delta \epsilon)$ for an imposed strain rate $\dot{\epsilon}_a=10^{-5}$ and driving spring stiffness $k=0.1$ for varying system sizes defined by the initial number of dislocations $N_0$. The upper cutoff evolution of the distributions are shown in the insets.}
	\label{fig:size}
\end{figure}

First we investigate the system size effects by varying the number of initial dislocations $N_0$, while fixing the imposed strain rate $\dot{\epsilon}_a=10^{-5}$ and the spring stiffness $k=0.1$. Fig.~\ref{fig:size} shows representative single-sample (dashed lines) and the ensemble-averaged (solid lines) stress-strain curves. Smaller systems results in accumulation of larger stress during the loading process, and the single-sample curves exhibit high deviations from the averaged curves. Additionally, the small number of dislocations allows the formation of fewer dislocation structures, resulting in abrupt single-sample stress-strain curves. Above a number of initial dislocations $N_0=900$ the system size dependence of the accumulated stress starts to be weaker, and the fluctuations around the average curves (i.e., the dislocation avalanches) become smaller.

The integrated distributions of the stress drop magnitudes $P_{\mathrm{INT}}(\Delta\sigma)$ and strain increments $P_{\mathrm{INT}}(\Delta\epsilon)$ for varying number of initial dislocations $N_0$ are shown in Figs.~\ref{fig:size}(b) and (c), respectively. The resulting cutoff scales of the distributions as a function of $N_0$ are shown in the corresponding insets. Generally, small systems result in larger cutoffs, however the evolution of the cutoff with increasing system size slows down to an extent that the model at hand with the system sizes reachable numerically is unable to reach the vanishing stress drop magnitudes and strain increments of macroscopic systems. This is due to the long range pair interaction of dislocations, implying that the computational time increases quadratically with the number of dislocations.
Thus, for the rest of this work the number of initial dislocations is held constant at $N_0=1600$, corresponding to a linear system size of $L=200$.

\subsection{\label{subsec:3.1} Stress-strain curves}

\begin{figure}[t!]
	\centering
	\resizebox{0.95\columnwidth}{!}{\includegraphics{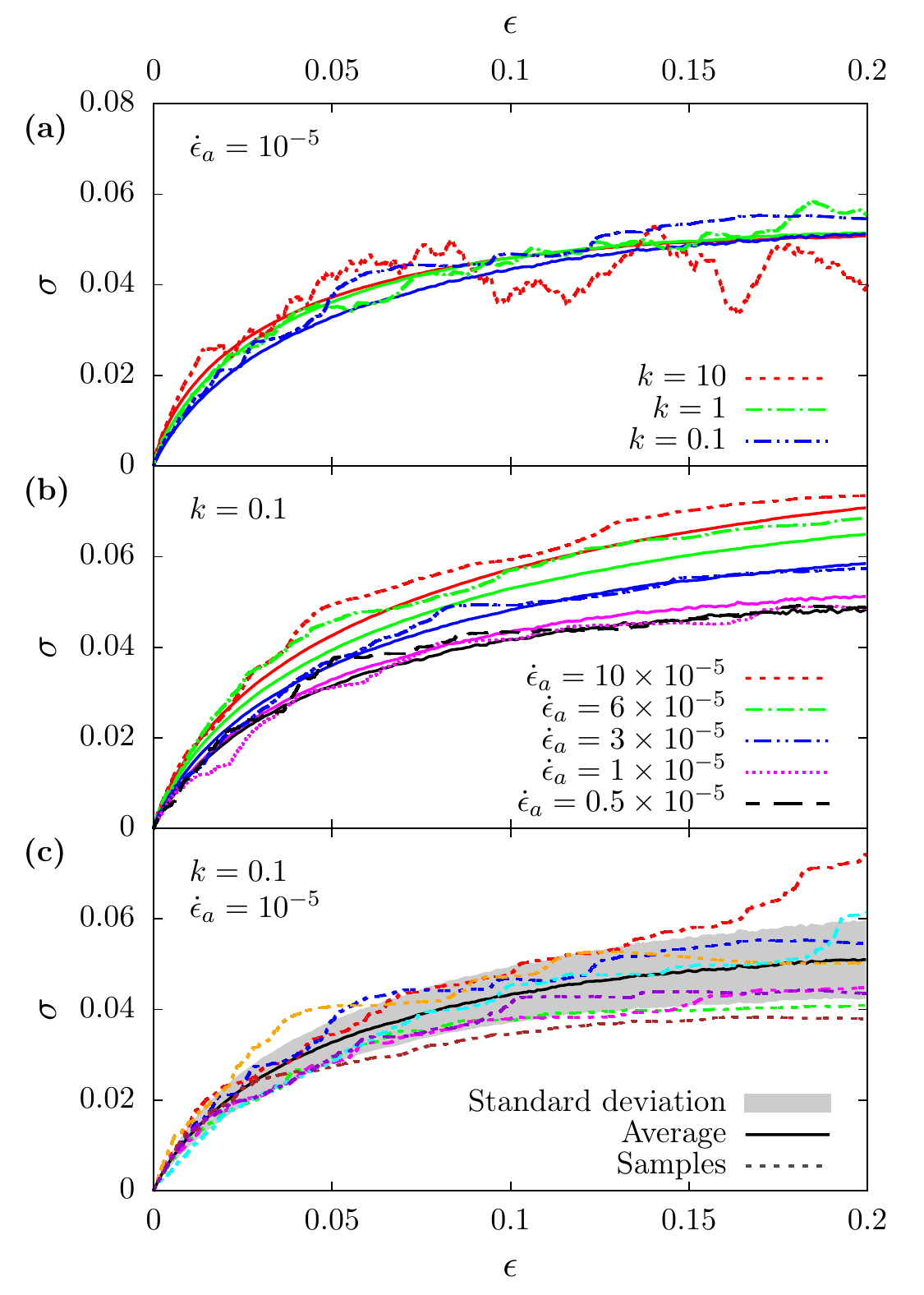}}
	\resizebox{0.95\columnwidth}{!}{\includegraphics{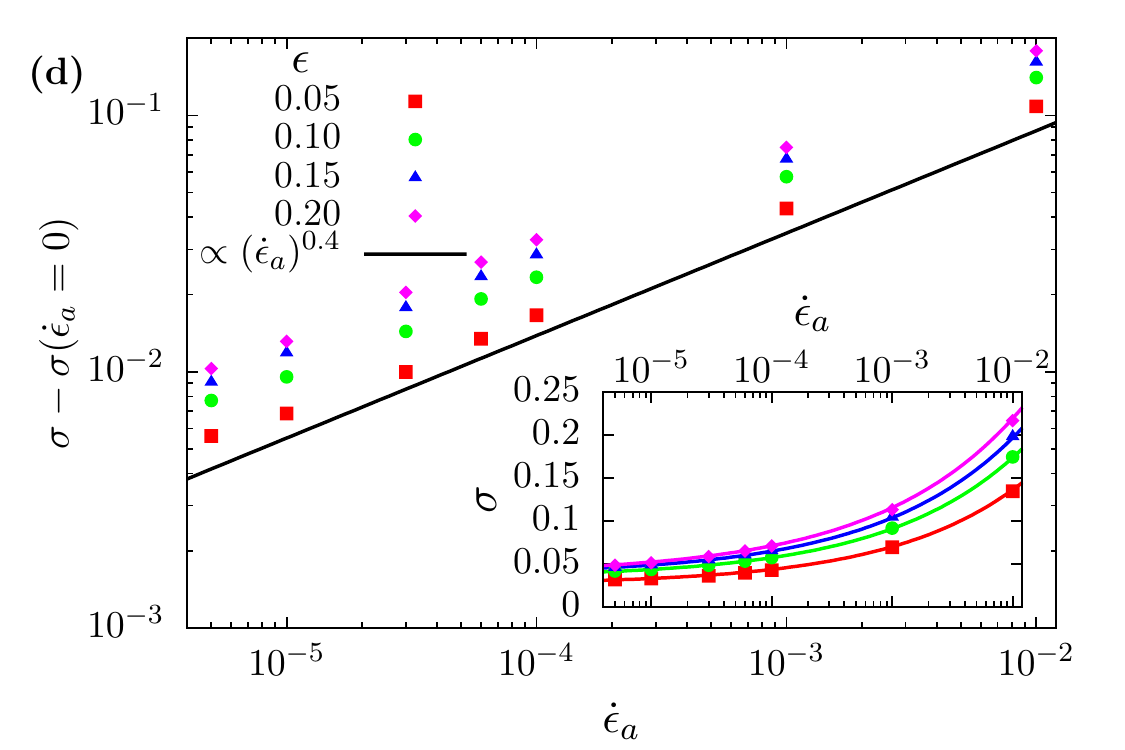}}
	\caption{Average stress-strain curves $\langle \sigma(\epsilon) \rangle$ (with the average taken over different realizations of the initial configuration, solid lines) and examples of individual, single-sample stress-strain curves (patterned lines) for (a) an imposed strain rate $\dot{\epsilon}_a=10^{-5}$ and different driving spring stiffnesses $k$; (b) a fixed driving spring stiffness $k=0.1$ and different imposed strain rates $\dot{\epsilon}_a$ (the corresponding average curve increases with increasing $\dot{\epsilon}_a$); (c) both fixed driving spring stiffness $k=0.1$ and strain rate $\dot{\epsilon}_a=10^{-5}$, showing additionally the standard deviation (shaded region) and multiple single-sample stress-strain curves. (d) The dependence of the ensemble-averaged stress-strain curves on the strain rate $\dot{\epsilon}_a$ at specific strain values $\epsilon$ obeying the shifted power law [Eq. \eqref{eq:4}].}
	\label{fig:2}
\end{figure}

Examples of single-sample as well as ensemble-averaged stress-strain curves are shown in Figs.~\ref{fig:2}(a), (b) and (c) with dashed and solid lines, respectively. In general, the properties of these curves depend on the driving parameters $\dot{\epsilon}_a$ and $k$. Varying $k$ with a fixed $\dot{\epsilon}_a$ [see Fig.~\ref{fig:2}(a)] shows that even if the average stress-strain curves are largely unaffected by the value of $k$, the magnitude of the fluctuations around the average curve increases significantly with increasing $k$. On the other hand, varying $\dot{\epsilon}_a$ with a fixed $k$ [see Fig.~\ref{fig:2}(b)] results in a significant dependence of the average stress-strain curves on $\dot{\epsilon}_a$, while the fluctuations visible in the single-sample stress-strain curves appear to be less sensitive to $\dot{\epsilon}_a$. In order to visualize the sample-to-sample variation of the individual stress-strain curves Fig.~\ref{fig:2}(c) shows multiple single-sample curves along with the ensemble-averaged stress-strain curve and its standard deviation for the case with $\dot{\epsilon}_a=10^{-5}$ and $k=0.1$.

A detailed study of the properties of the average stress-strain curves for different imposed strain rates is shown in Fig.~\ref{fig:2}(d). The dependence of the accumulated average stress at a given strain $\epsilon$ on the imposed strain rate $\dot{\epsilon}_a$ is well-described by a shifted power law
\begin{equation}\label{eq:4}
\sigma(\epsilon,\dot{\epsilon}_a) = \sigma(\epsilon,\dot{\epsilon}_a=0)+A\dot{\epsilon}_a^b,
\end{equation}
with the exponent $b \approx 0.4$ and the shift equal to the stress at zero strain rate. The exponent $b$ characterizing the rate dependence of the stress level at a given strain in our 2D DDD simulations is smaller than that in the linear dependence found recently for 3D simulations in Ref.~\cite{fan2021strain}. This applies to all the tested strains during the loading process.

\subsection{\label{subsec:3.2} Stress drop and strain burst statistics}

\begin{figure}[t!]
	\centering
	\resizebox{0.95\columnwidth}{!}{\includegraphics{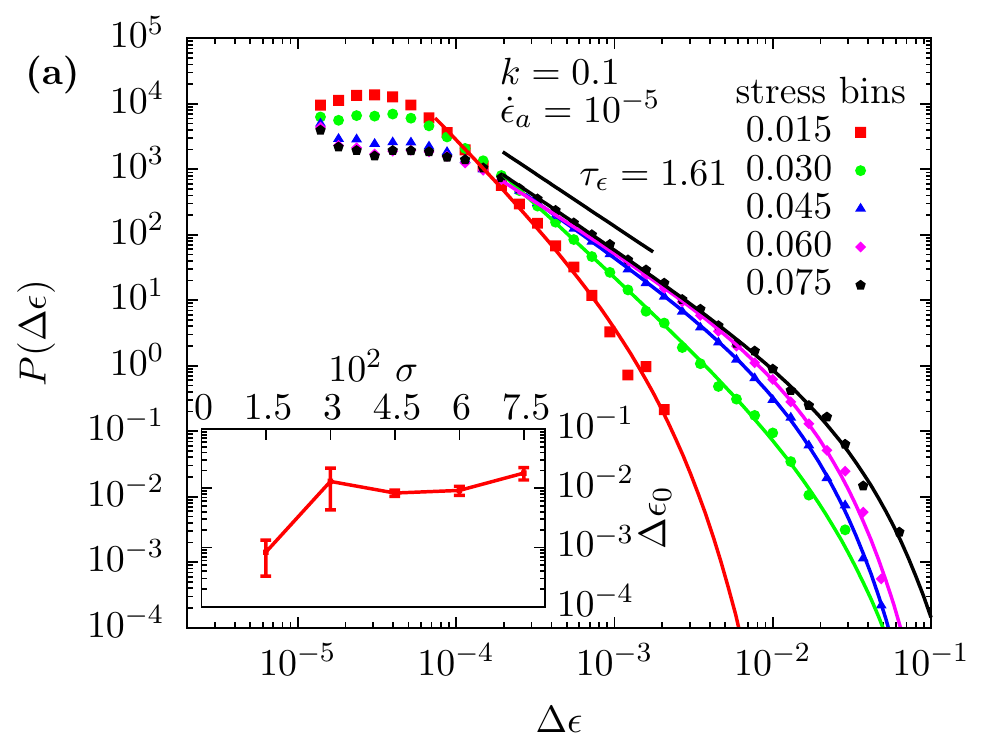}}
	\resizebox{0.95\columnwidth}{!}{\includegraphics{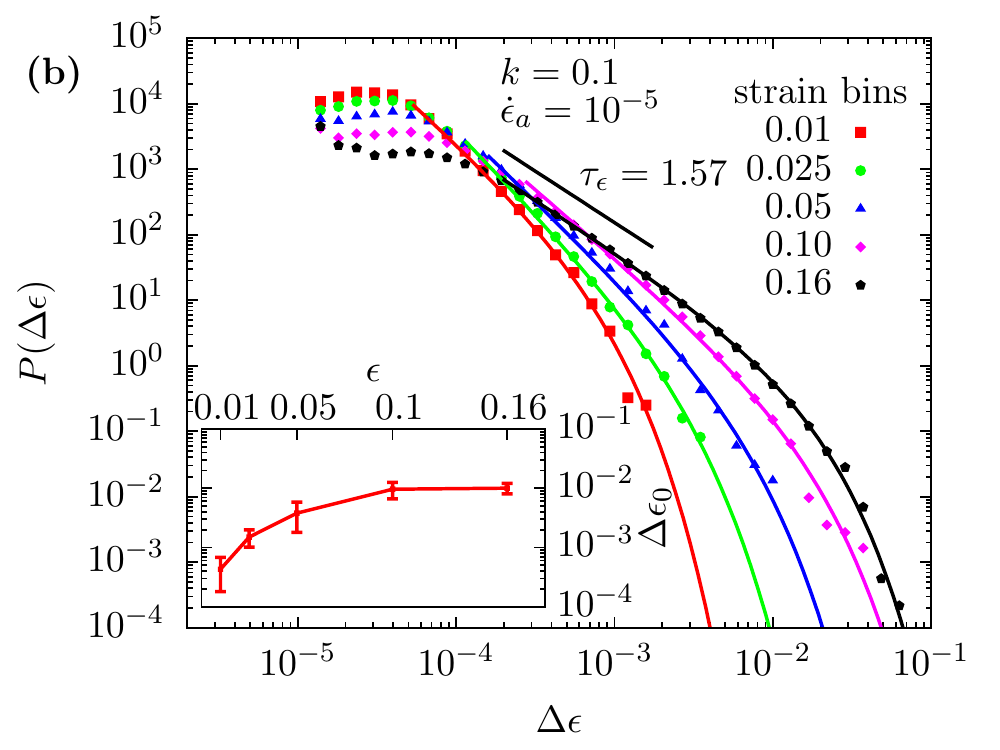}}
	\caption{(a) Stress and (b) strain binned distributions of the strain increments $P(\Delta \epsilon)$ for an imposed strain rate $\dot{\epsilon}_a=10^{-5}$ and driving spring stiffness $k=0.1$. The legend shows the upper limit of the corresponding stress/strain bins. The insets show the evolution of the upper cutoff with the stress and strain levels, respectively.}
	\label{fig:3}
\end{figure}

First, we investigate the non-stationary behavior of the distributions using the curves obtained for the imposed strain rate $\dot{\epsilon}_a=10^{-5}$ and spring stiffness $k=0.1$. Fig.~\ref{fig:3} shows the strain burst distributions $P(\Delta \epsilon)$ with stress [Fig.~\ref{fig:3}(a)] and strain [Fig.~\ref{fig:3}(b)] binning, where the values in the legend represent the upper limits of the bins. A strain burst is characterized by the stress and strain values at which the event starts. To assure that every strain burst finishes before the simulation ends at $\epsilon=0.2$ a strain interval above the largest bin is not included. The stress-binned distributions can be well-described by power laws with exponential cutoffs  [Eq.~\eqref{eq:powerlaw}]. The data to be fitted was chosen so that the fitting error in the pure power law regime would be minimal, and the fits of this form are indicated by lines in Fig.~\ref{fig:3}. The cutoffs, shown in the insets of Fig.~\ref{fig:3}, exhibit a dependence on stress and strain. The general trend is that for small stresses and strains, the cutoffs increase with stress and strain (i.e., the avalanche dynamics is initially non-stationary), but the growth of the cutoffs seems to slow down and saturate for larger stresses/strains, resulting in an approach to a quasi-stationary avalanche dynamics towards the end of the simulation. For the largest bins considered, both stress and strain binned distributions appear to exhibit a power law exponent $\tau_\epsilon \approx 1.6$, i.e., a significantly larger value than found in previous studies employing stress-controlled loading~\cite{ispanovity2014avalanches}. In what follows, due to this quasi-stationary nature of the avalanche dynamics after the initial transient, we focus on considering the "integrated" distributions, where all events irrespective of the stress or strain at which they occur are included in the same distribution.

\begin{figure}[t]
	\centering
	\resizebox{0.95\columnwidth}{!}{\includegraphics{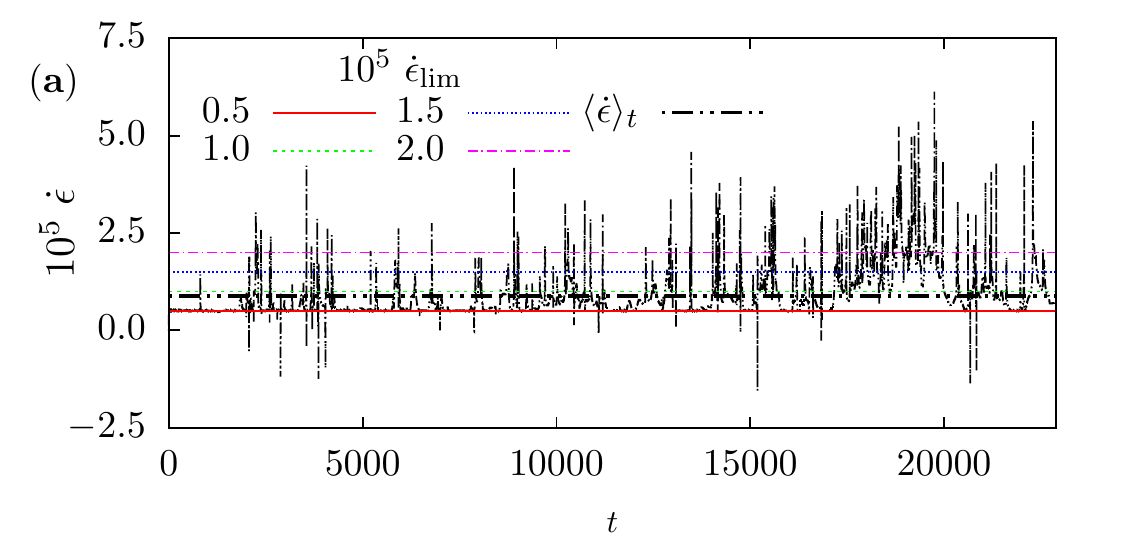}}
		\resizebox{0.95\columnwidth}{!}{\includegraphics{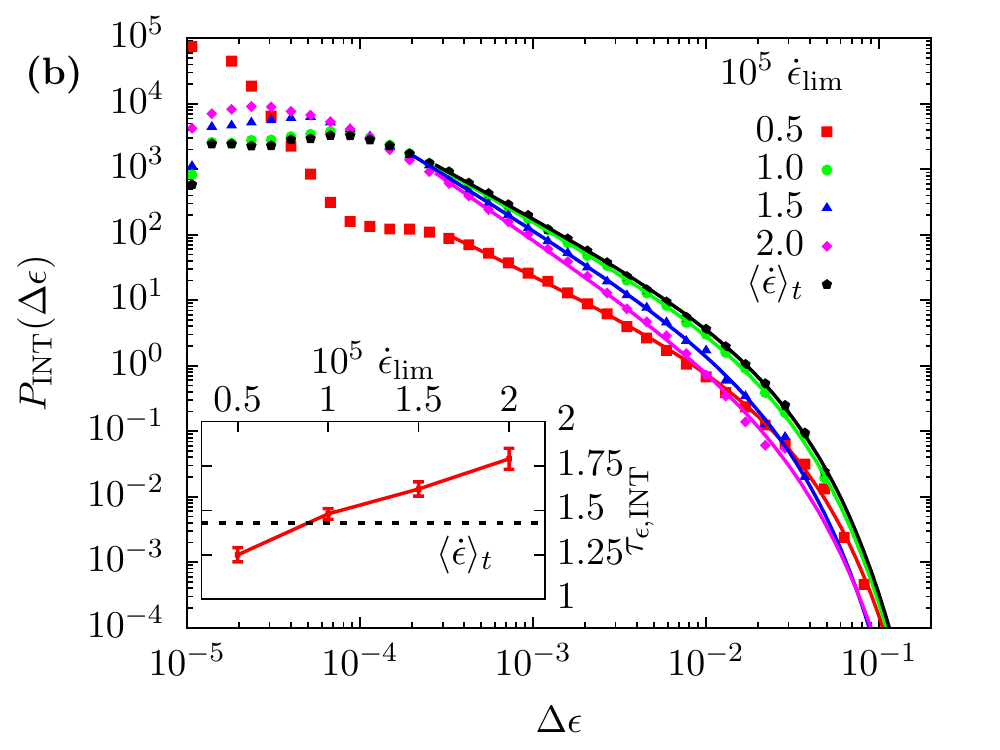}}
	\caption{(a) Sample strain rate $\dot{\epsilon}$ signal as a function of time for stress-controlled loading including different avalanche thresholds $\dot{\epsilon}_{\mathrm{lim}}$ (with $0.5 \times 10^{-5}$ being the driving threshold) and the strain rate time average $\langle \dot{\epsilon} \rangle_t$; and (b) strain increment distributions of samples with stress-controlled loading for varying avalanche thresholding, including the strain rate time average $\langle \dot{\epsilon} \rangle_t$ as a threshold and the resulting power law exponents in the inset.}
	\label{fig:stressLoad}
\end{figure}

Here, we aim at shedding some light on the origin of the large exponent value $\tau_\epsilon \approx 1.6$ found above. In previous studies using quasistatic stress-controlled loading a collective velocity threshold was driving the loading process~\cite{ispanovity2014avalanches,ovaska2015quenched,lehtinen2016glassy,salmenjoki2020plastic}, meaning that a collective velocity below the threshold resulted in stress increasing with a small rate, while during a collective velocity above the threshold strain was accumulated at a constant stress. The same collective velocity threshold also defined the avalanches. A similar protocol, defining the loading process by a strain rate threshold, leads to staircase-like stress-strain curves, where we take the avalanche definitions (in contrast to the imposed strain rate $\dot{\epsilon}_a$ of the strain-controlled loading) to not follow directly from the loading protocol, and instead consider an arbitrary avalanche threshold $\dot{\epsilon}_\mathrm{lim}$. Fig.~\ref{fig:stressLoad}(a) shows a sample strain rate signal for a strain rate threshold based quasistatic stress-controlled loading, where the horizontal lines (including the employed driving threshold $\dot{\epsilon}_{\mathrm{lim}}=0.5\times10^{-5}$ and the time average of the strain rate $\langle \dot{\epsilon} \rangle_t$) represent possible avalanche thresholds. Notice especially that with the exception of a few short negative spikes, $\dot{\epsilon}(t)$ is either above the driving threshold, or only very slightly below it, implying that the mean value of the signal is well above the driving threshold. The integrated strain burst distributions $P_{\mathrm{INT}}(\Delta\epsilon)$ for these avalanche thresholds (accounting for the variation of the strain rate time average $\langle \dot{\epsilon} \rangle_t$ from sample to sample) are shown in Fig.~\ref{fig:stressLoad}(b). The tails of the distributions are fitted with a power law with exponential cutoff according to Eq.~(\ref{eq:powerlaw}), indicated by solid lines in the main panel and the corresponding exponents $\tau_{\epsilon,\mathrm{INT}}$ in the inset. For the lowest avalanche threshold, which corresponds to the driving threshold, the power law exponent $\tau_{\epsilon,\mathrm{INT}}=1.25\pm0.03$ is comparable to the literature value of $1.3$ (where collective velocity threshold was used as both the driving and the avalanche threshold)~\cite{ispanovity2014avalanches}. However, as the avalanche threshold increases and distances itself from the driving threshold, the power law exponent is found to increase monotonically. Notice that as the threshold increases, the avalanche events defined by the thresholding process are increasingly the individual short spikes in the rather spiky $\dot{\epsilon}(t)$ signal, and our analysis shows that these exhibit a larger, threshold-dependent exponent as compared to the events defined by the driving threshold. Given that strain-controlled loading imposes a threshold equal to the mean of the $\dot{\epsilon}(t)$ signal, this finding could explain the larger exponent found above for that driving protocol.

\begin{figure}[t!]
	\centering
	\resizebox{0.95\columnwidth}{!}{\includegraphics{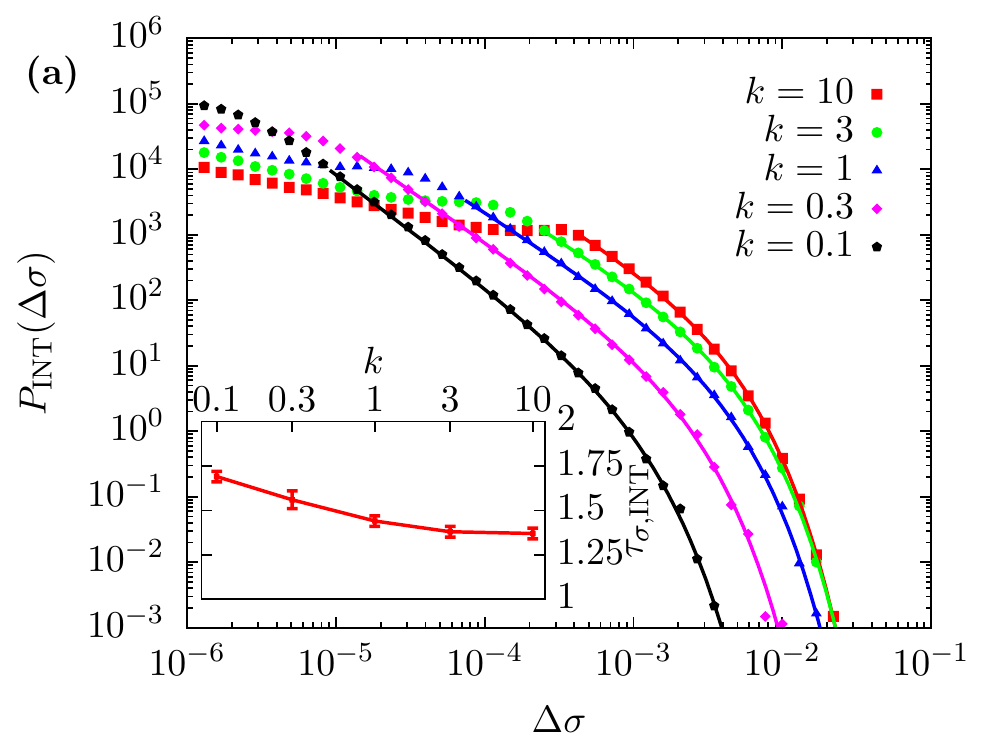}}
	\resizebox{0.95\columnwidth}{!}{\includegraphics{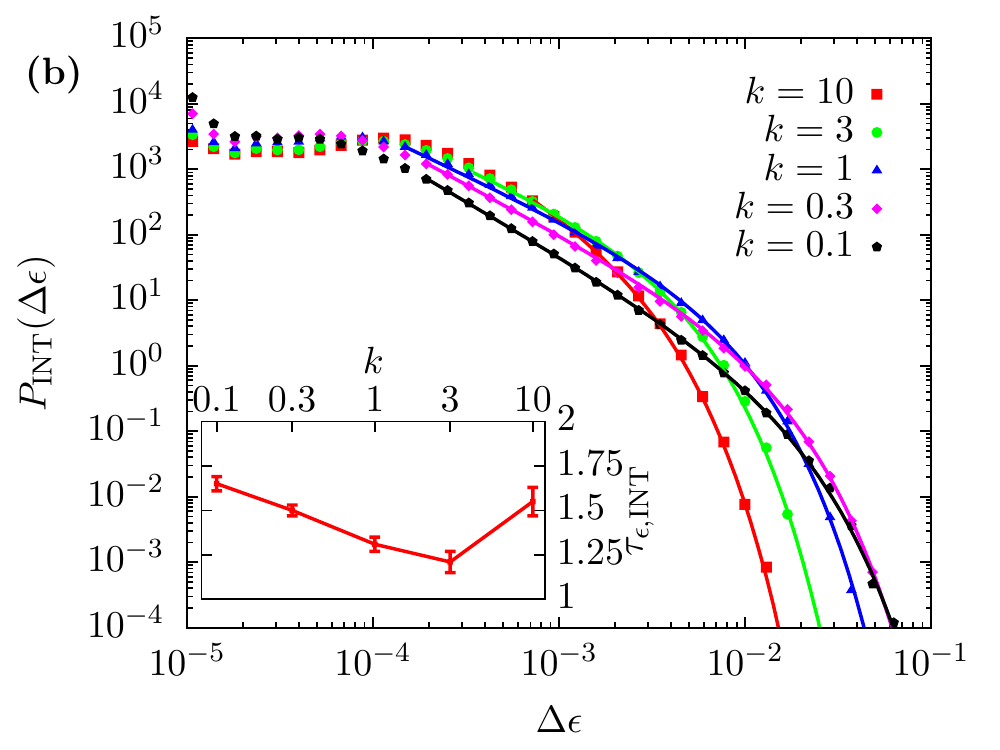}}
	\resizebox{0.95\columnwidth}{!}{\includegraphics{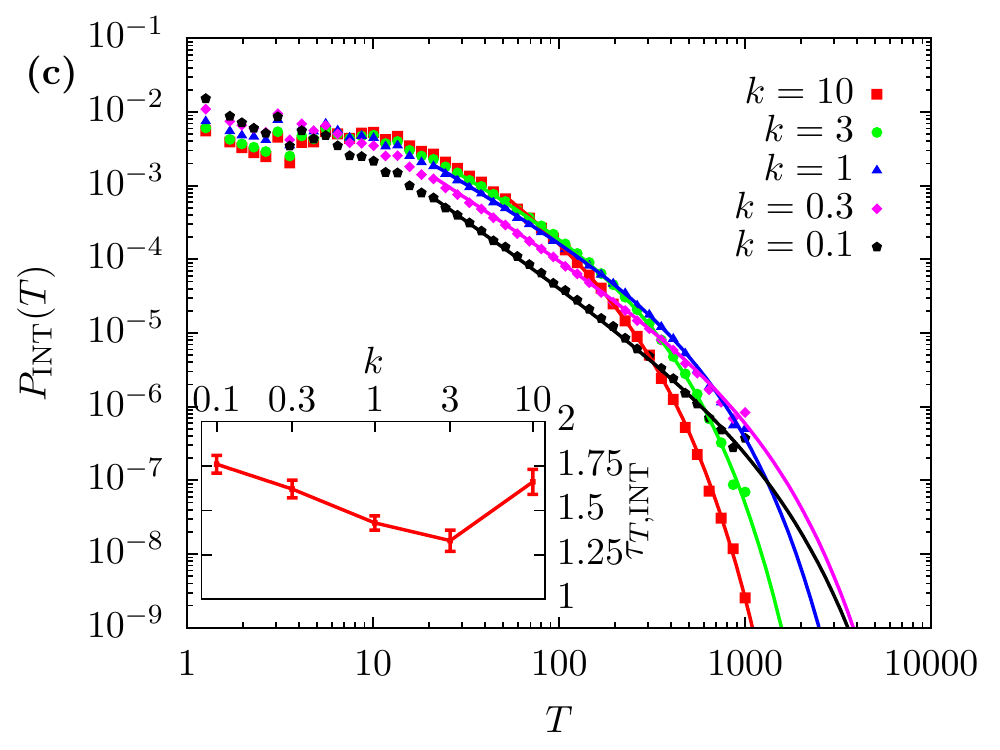}}
	\caption{Integrated distributions of the (a) stress drop magnitudes $P_{\mathrm{INT}}(\Delta \sigma)$, (b) strain increments $P_{\mathrm{INT}}(\Delta \epsilon)$ and (c) avalanche durations $P_{\mathrm{INT}}(T)$ for an imposed strain rate $\dot{\epsilon}_a=10^{-5}$ and varying driving spring stiffnesses $k$, with corresponding power law exponent obtained from the fits shown in the insets.}
	\label{fig:4}
\end{figure}

We then proceed to study the integrated dislocation avalanche distributions using strain-controlled loading. Fig.~\ref{fig:4} shows the integrated distributions of the stress drop magnitudes $P_{\mathrm{INT}}(\Delta \sigma)$ [Fig.~\ref{fig:4}(a)] from our strain-controlled simulations, those of the corresponding strain increments $P_{\mathrm{INT}}(\Delta \epsilon)$ [Fig.~\ref{fig:4}(b)], as well as of the avalanche durations $P_{\mathrm{INT}}(T)$ [Fig.~\ref{fig:4}(c)] for an imposed strain rate of $\dot{\epsilon}_a=10^{-5}$ and spring stiffnesses $k$ varying in a range of two decades (corresponding to different levels of strain control). The tails of the distributions are again fitted with a power law terminated at an exponential cutoff according to Eq.~\eqref{eq:powerlaw}. In the case of the stress drop distributions [Fig.~\ref{fig:4}(a)], the fitting range depends heavily on the spring stiffness $k$, such that the range of $\Delta \sigma$-values following a power law distribution extends to smaller $\Delta \sigma$-values for smaller $k$. Below this interval the distributions flatten and hence can no longer be described by a power law. The resulting power law exponents $\tau_{\sigma,\mathrm{INT}}$ as a function of $k$ lie within the interval $[1.3-1.7]$, such that the largest exponent value $\tau_{\sigma,\mathrm{INT}} \approx 1.7$ is obtained for the smallest $k$.  [see the inset of Fig~\ref{fig:4}(a)]. This small-$k$ limit also has the longest power law scaling regime, while the largest $k$-values considered lead to a very narrow (possibly non-existent for $k=10$) scaling regime, and hence the exponent values from the fits in this limit might not be very accurate (the same applies to the $k=10$ strain increment and event duration distributions, discussed below). As a consequence, the large $k$ limit is expected to break scale invariance. The cutoff scale $\Delta\sigma_0$ of the stress drop distribution is found to increase with increasing $k$.

The strain increment distributions $P_{\mathrm{INT}}(\Delta \epsilon)$ [Fig.~\ref{fig:4}(b)] exhibit many similarities to the stress drop distributions, such as flattening of the distributions below the pure power law regime and power law exponent $\tau_{\epsilon,\mathrm{INT}}$ depending on $k$ in a similar way. Specifically, the longest power law regime is again obtained for the small $k=0.1$, and is characterized by an exponent $\tau_\epsilon \approx 1.7$. However, the lower limit of the power law regime is found to be roughly independent of $k$, and the cutoff scale $\Delta\epsilon_0$ exhibits the opposite dependence on $k$ compared to the stress drop distributions, i.e., a larger $\Delta\epsilon_0$ is observed for smaller $k$. The similarities between the stress drop magnitude $\Delta\sigma$ and strain increment $\Delta\epsilon$ distributions can be interpreted by considering the driving equation Eq.~\eqref{eq:drive}, ending up with the expression $\Delta \sigma= k\left[\dot{\epsilon}_a T-\Delta \epsilon\right]$, where $T$ is the avalanche duration. Thus, in the ideal case of an infinitesimally low applied strain rate $\dot{\epsilon}_a$, the magnitudes of these are simply proportional, $\Delta \sigma = -k\Delta \epsilon$, so that one expects the stress drop and strain burst distributions to scale with the same exponent. Finite $\dot{\epsilon}_a$ results in a correction such that $\Delta \sigma$ is a sum of $k\dot{\epsilon}_a T$ and $-k\Delta\epsilon$, with both of these contributions obeying their own truncated power law distributions.

The avalanche duration distributions $P_\mathrm{INT}(T)$ [Fig.~\ref{fig:4}(c)] show a very similar behavior to the strain increment distributions, with the evolution of the cutoff scale with $k$ and the lower limit of the fitting range being independent of $k$ being the common properties. Also the $k$-dependent power law exponents $\tau_{T,\mathrm{INT}}$ obtained from the fits are rather similar to the corresponding $\tau_{\epsilon,\mathrm{INT}}$'s, with $\tau_{T,\mathrm{INT}} \approx 1.75$ for the small $k=0.1$.

\begin{figure}[t!]
	\centering
	\resizebox{0.95\columnwidth}{!}{\includegraphics{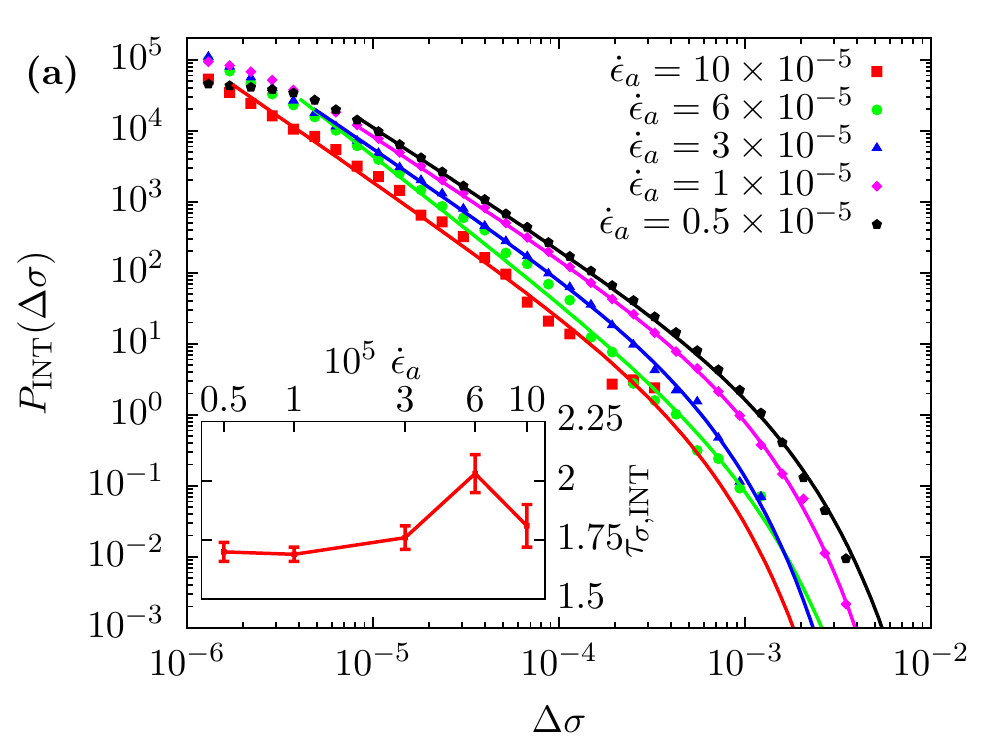}}
	\resizebox{0.95\columnwidth}{!}{\includegraphics{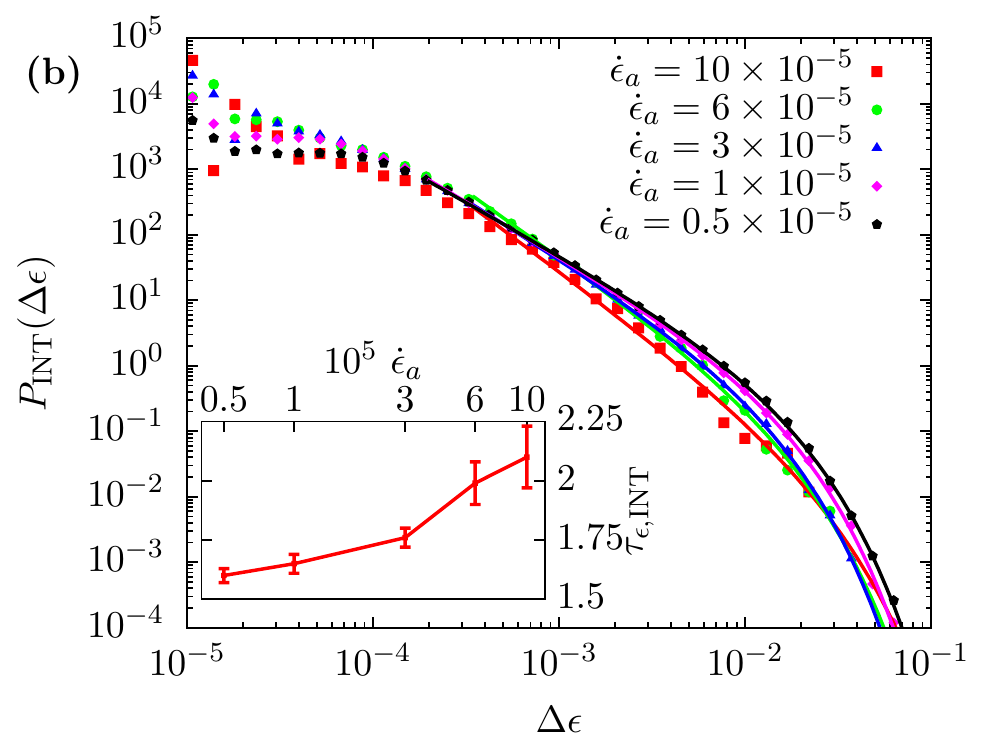}}
	\resizebox{0.95\columnwidth}{!}{\includegraphics{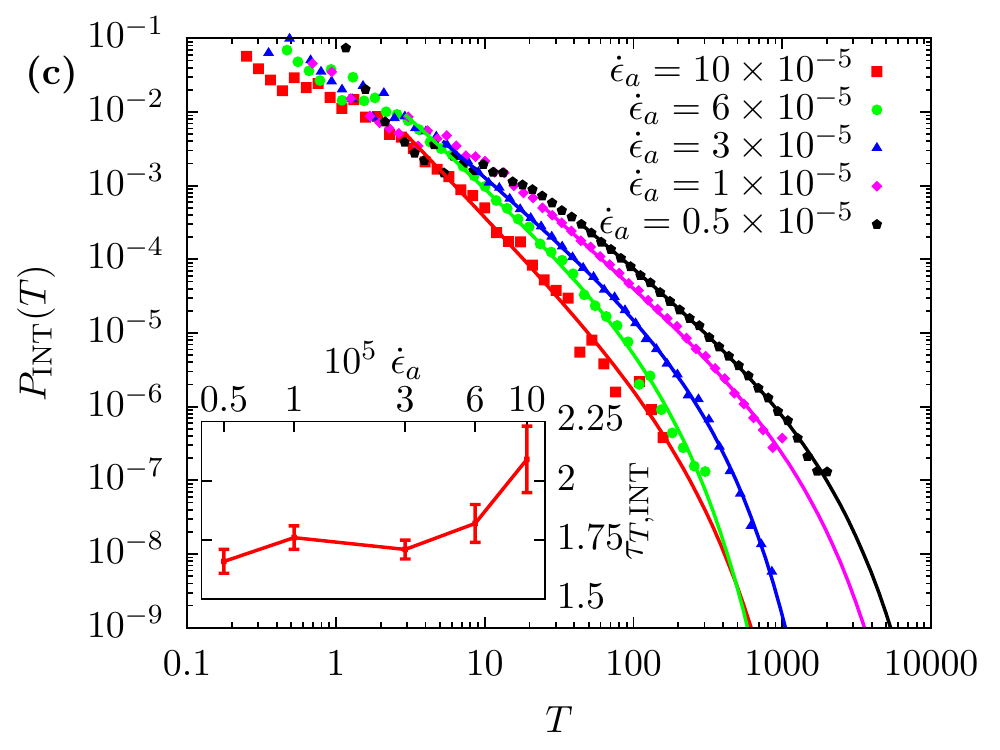}}
	\caption{Integrated distributions of the (a) stress drop magnitudes $P_{\mathrm{INT}}(\Delta \sigma)$, (b) strain increments $P_{\mathrm{INT}}(\Delta \epsilon)$ and (c) avalanche durations $P_{\mathrm{INT}}(T)$ for a fixed driving spring stiffness $k=0.1$ and varying imposed strain rates $\dot{\epsilon}_a$, with corresponding power law exponent obtained from the fits shown in the insets.}
	\label{fig:5}
\end{figure}

Fig.~\ref{fig:5} shows the corresponding distributions for a fixed spring stiffness $k=0.1$ ("soft spring"), varying the imposed strain rate $\dot{\epsilon}_a$. First, the stress drop distributions $P_\mathrm{INT}(\Delta \sigma)$ in Fig.~\ref{fig:5}(a) have a cutoff $\Delta \sigma_0$ shifting to larger values as the imposed strain rate $\dot{\epsilon}_a$ decreases. The combination of a soft spring ($k=0.1$) and high strain rate ($\dot{\epsilon}_a = 10 \times 10^{-5}$) results in a very small number of stress drop/strain burst events per stress-strain curve, and hence the distribution cutoff for $\dot{\epsilon}_a = 10 \times 10^{-5}$ is not well-resolved. The power law part is most pronounced in the limit of low strain rates, with the exponent for $\dot{\epsilon}_a = 0.5 \times 10^{-5}$ given by $\tau_{\sigma,\mathrm{INT}} \approx 1.7$.

Fig.~\ref{fig:5}(b) shows the strain rate dependence of the strain increment (strain burst) distribution $P_{\mathrm{INT}}(\Delta \epsilon)$. In the limit of low strain rate, the strain-controlled distributions have an exponent $\tau_{\epsilon,\mathrm{INT}} \approx 1.6$ and $\tau_{\epsilon,\mathrm{INT}}$ seems to slightly increase with increasing $\dot{\epsilon}_a$. Notice however that for large $\dot{\epsilon}_a$ there are not that many events per stress-strain curve, and hence the accuracy of the measured exponent is limited.

The avalanche duration distributions $P_\mathrm{INT}(T)$ for the fixed spring stiffness [Fig.~\ref{fig:5}(c)] visualizes well how the events take place for varying strain rates $\dot{\epsilon}_a$. The lowest $\dot{\epsilon}_a$ generally results in longer events, and the cutoff $T_0$ of the event duration distribution decreases with increasing $\dot{\epsilon}_a$. The power law exponent $\tau_{T,\mathrm{INT}}$ is found to be around 1.7 for the lowest strain rate, and possibly exhibit a modest increase with $\dot{\epsilon}_a$. However, the lack of statistics for the highest strain rates, together with the limited extent of the power law scaling regime, might render the high-$\dot{\epsilon}_a$ exponents somewhat inaccurate.

Having seen the behavior of the integrated distributions for varying parameters $\dot{\epsilon}_a$ and $k$, we come to conclusions relating the present model to the previously implemented systems using quasistatic stress-controlled loading. In the limit $k\rightarrow 0$ and fixed $\dot{\epsilon}_a$ our model would approach stress-controlled loading, however the threshold value applied to the strain rate signal to define the avalanches needs to be considered additionally. Thus, enforcing a very low imposed strain rate $\dot{\epsilon}_a$, so that the threshold (equalling $\dot{\epsilon}_a$) would approach zero, is also required in order to reproduce stress-controlled simulations. The difficulty arises from the implementation of the strain-controlled loading, where taking the limit of both $\dot{\epsilon}_a$ and $k$ going to zero would result in infinitely slowly progressing simulations. Also, the avalanche distributions even for the smallest $\dot{\epsilon}_a$ and $k$ considered in our study appear to exhibit differences compared to stress-controlled simulations (notably,  $\tau_{\epsilon,\mathrm{INT}}\approx 1.3$ for stress-controlled loading~\cite{ispanovity2014avalanches}). Hence, using our simulation results, we cannot conclude that our avalanche behavior would be fully in agreement with the stress-controlled loading mode, even if in the limit of $k\rightarrow0$ one would expect to reach a limit characteristic of quasistatic stress-controlled loading.

\begin{figure}[t!]
	\centering
	\resizebox{0.95\columnwidth}{!}{\includegraphics{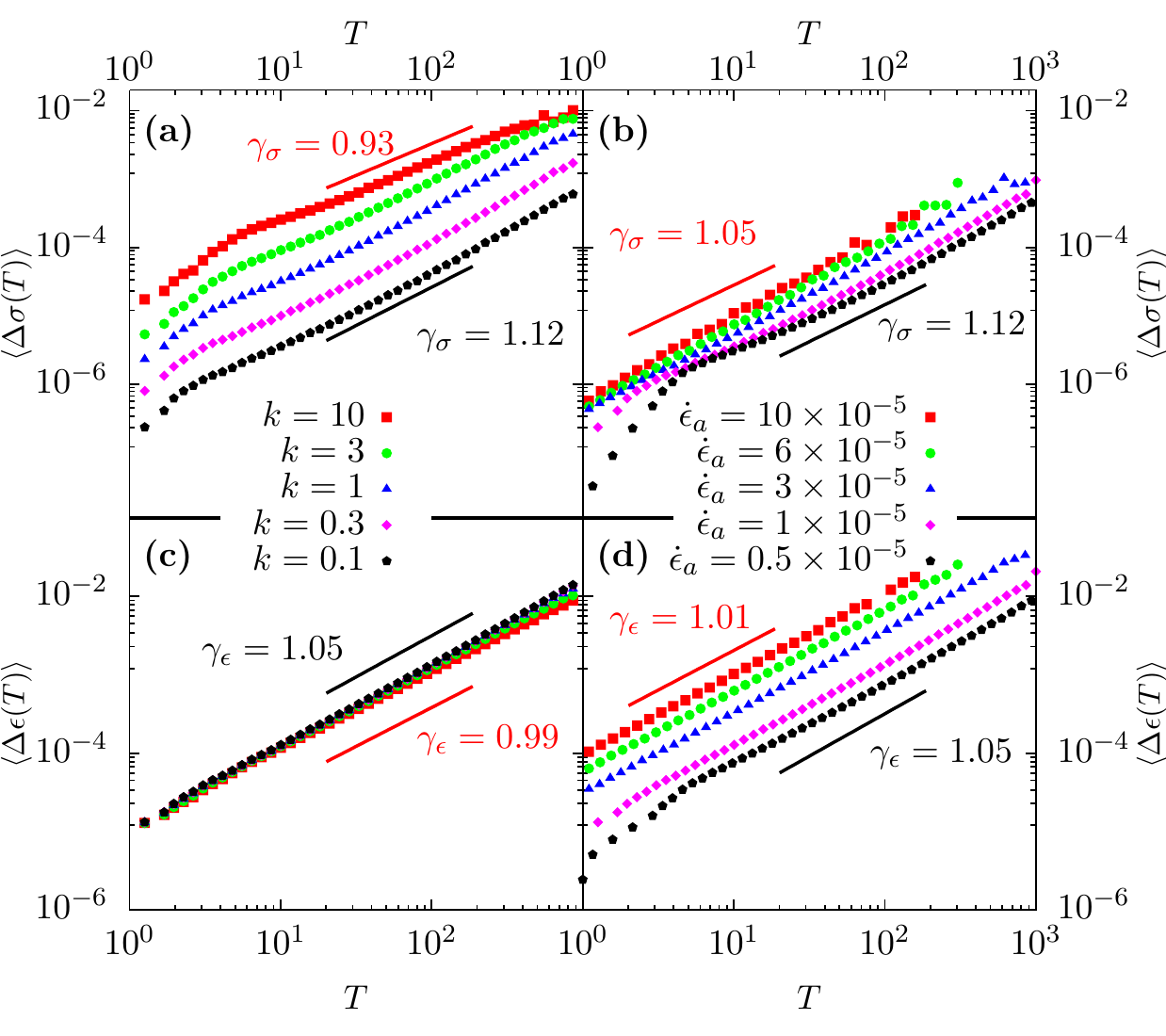}}
	\caption{Scaling of the (a),(b) average stress drop magnitude $\langle \Delta \sigma (T) \rangle$ and (c),(d) average strain increment $\langle \Delta \epsilon (T) \rangle$ with the event duration $T$; (a),(c) for an imposed strain rate $\dot{\epsilon}_a=10^{-5}$ and varying driving spring stiffnesses $k$, and (b),(d) for a fixed driving spring stiffness $k=0.1$ and varying imposed strain rates $\dot{\epsilon}_a$, respectively. 
	Lines correspond to power law fits to the data with largest and smallest $k$ and $\dot{\epsilon}_a$, with the corresponding exponent values indicated in the plot.}
	\label{fig:6}
\end{figure}

The scaling of the average stress drop magnitude $\langle \Delta \sigma (T) \rangle$ and average strain increment $\langle \Delta \epsilon (T) \rangle$ with the event duration $T$ is reported in Fig.~\ref{fig:6}. In the mid-parts of their duration ranges, both scale with the event duration as $T^{\gamma_x}$, where $x=\sigma$ and $\epsilon$ for stress drops and strain increments, respectively. The exponents $\gamma_\sigma$ [Figs.~\ref{fig:6}(a) and (b)] and $\gamma_\epsilon$ [Figs.~\ref{fig:6}(c) and (d)] are both close to unity -- in good agreement with the scaling relation $\gamma_x = (\tau_T-1)/(\tau_x-1)$, given that $\tau_x$ and $\tau_T$ were above found to exhibit similar values -- but possibly exhibit a weak dependence on $k$ and $\dot{\epsilon}$. Solid lines in Fig.~\ref{fig:6} indicate the exponent values in the limiting cases, i.e., for the largest and smallest $k$ and $\dot{\epsilon}_a$. In the limit of low strain rate and soft spring ($\dot{\epsilon}_a = 0.5 \times 10^{-5}$ and $k=0.1$), we find $\gamma_\sigma \approx 1.1$ and $\gamma_\epsilon \approx 1.05$. We note that the value of the scaling exponent obtained for quasistatic stress-controlled loading, $\gamma_\epsilon=1.32$ \cite{ispanovity2014avalanches}, is significantly higher than our $\gamma_{\epsilon}$ for strain-controlled loading, again highlighting the difference between the two loading protocols. It is also worth pointing out that the amplitudes of the $\langle \Delta \sigma (T) \rangle$ and $\langle \Delta \epsilon (T) \rangle$ curves depend on $k$ and $\dot{\epsilon}_a$. The amplitude of $\langle \Delta \sigma (T) \rangle$ increases quite strongly with $k$, while that of $\langle \Delta \epsilon (T) \rangle$ is essentially independent of $k$. At the same time, amplitudes of both $\langle \Delta \sigma (T) \rangle$ and $\langle \Delta \epsilon (T) \rangle$ increase with $\dot{\epsilon}_a$, with the amplitude of $\langle \Delta \epsilon (T) \rangle$ exhibiting a stronger $\dot{\epsilon}_a$ dependence.

\subsection{\label{subsec:3.3} Average avalanche shapes}

In addition to the probability distributions of sizes and durations, as well as the scaling of the average size with the duration, avalanches are often characterized by their average shapes \cite{Papanikolaou2011,laurson2013avalancheshape}. In plasticity the event shape means the average strain rate profile from the start of the strain burst to the end of it, with the threshold strain rate $\dot{\epsilon}_a$ subtracted from the signal i.e., $\langle \dot{\epsilon} \left(\frac{t}{T}\right)-\dot{\epsilon}_a\rangle$. Following Ref.~\cite{laurson2013avalancheshape}, we expect such average shapes to be parameterized by the exponent $\gamma$, which we expect to equal $\gamma_\sigma$ characterizing the average stress drop magnitude scaling $\langle \Delta \sigma (T) \rangle$ with the event duration $T$ [given that $\Delta \sigma = k\int_0^T (\dot{\epsilon}-\dot{\epsilon}_a)\mathrm{d}t$ is proportional to the strain burst size obtained when subtracting the threshold strain rate $\dot{\epsilon}_a$ from the $\dot{\epsilon}(t)$ signal], and a parameter $a$ describing the temporal asymmetry of the avalanches,
\begin{eqnarray}\label{eq:2}
{\textstyle \langle \dot{\epsilon} \left(\frac{t}{T}\right)-\dot{\epsilon}_a\rangle \propto T^{\gamma-1}\left[ \frac{t}{T}\left(1-\frac{t}{T}\right) \right]^{\gamma-1} \left[ 1-a\left(\frac{t}{T}-\frac{1}{2}\right) \right] }.
\end{eqnarray}
Eq.~\ref{eq:2} consists of a symmetrical part parameterized by $\gamma$, and a lowest order correction to describe a weak asymmetry of the shape, quantified by $a$. 

\begin{table}[t]
	\caption{Minimum and maximum avalanche durations $T$ belonging to the power law part of $P_{\mathrm{INT}}(T)$ for different values of $k$ and $\dot{\epsilon}_a$, corresponding to the duration range included in the average when computing the average strain burst shapes shown in Fig.~\ref{fig:7}.}
	\label{tab:table1}
	\begin{tabular}{p{0.1\columnwidth}>{\centering}p{0.15\columnwidth}>{\centering}p{0.25\columnwidth}>{\centering\arraybackslash}p{0.2\columnwidth}}
		$k$   & $\dot{\epsilon}_a\ (\times 10^{-5})$ & $T_\mathrm{min}$ & $T_\mathrm{max}$ \\ \hline
		$3$   & $1$                                  & $28.41$          & $108.0$          \\
		$1$   & $1$                                  & $21.12$          & $145.3$          \\
		$0.3$ & $1$                                  & $21.12$          & $145.3$          \\
		$0.1$ & $1$                                  & $21.12$          & $195.5$          \\ \hline
		$0.1$ & $10$                                 & $2.75$           & $36.31$          \\
		$0.1$ & $6$                                  & $2.57$           & $39.30$          \\
		$0.1$ & $3$                                  & $4.73$           & $63.42$          \\
		$0.1$ & $0.5$                                & $38.41$          & $277.2$         
	\end{tabular}
\end{table}

\begin{figure}[t!]
	\centering
	\resizebox{0.95\columnwidth}{!}{\includegraphics{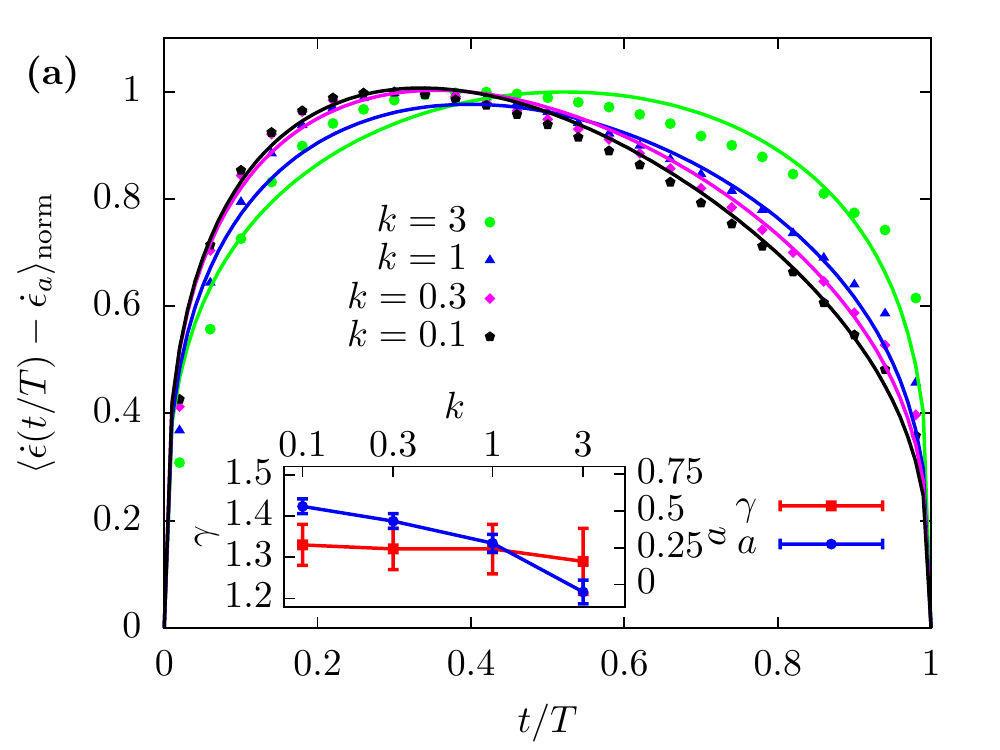}}
	\resizebox{0.95\columnwidth}{!}{\includegraphics{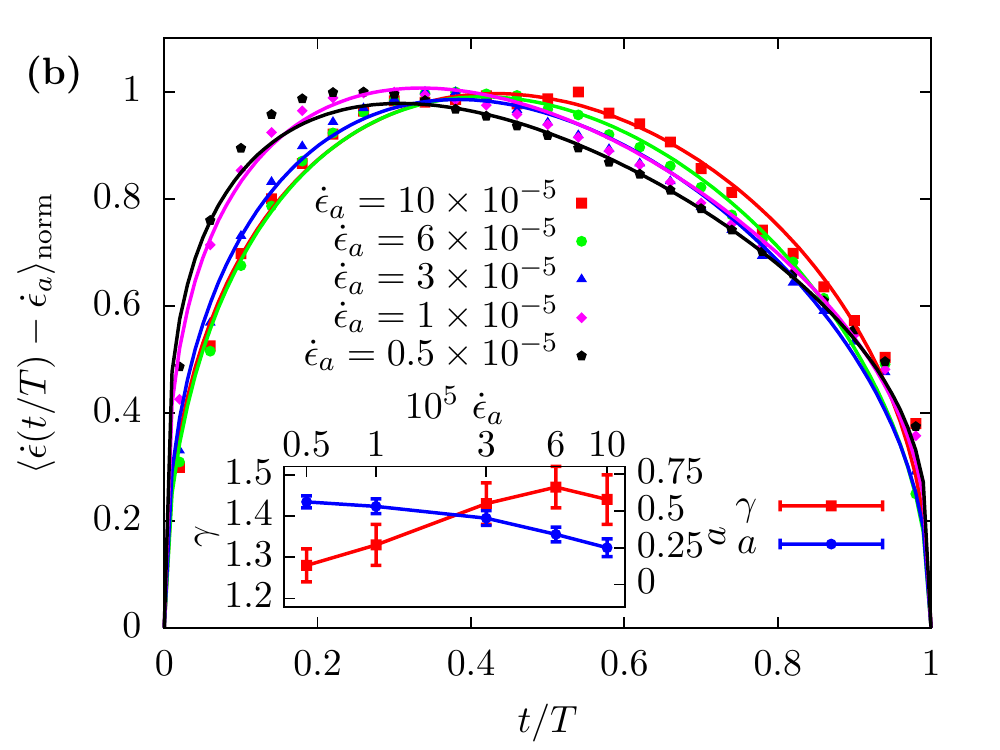}}
	\caption{Average strain burst (avalanche) shapes normalized by their maximum values $\langle \dot{\epsilon} (t/T) - \dot{\epsilon}_a\rangle_{\mathrm{norm}}$ for (a) an imposed strain rate $\dot{\epsilon}_a=10^{-5}$ and varying driving spring stiffnesses $k$; (b) a fixed driving spring stiffness $k=0.1$ and different imposed strain rates $\dot{\epsilon}_a$, also showing the resulting fitting parameters from equation \eqref{eq:2} in the insets.}
	\label{fig:7}
\end{figure}

Fig.~\ref{fig:7} shows the average strain burst shapes for events with avalanche durations $T$ found in the power law regime of the respective distributions, shown in Figs.~\ref{fig:4}(c) and \ref{fig:5}(c). Table~\ref{tab:table1} shows the minimum and maximum event durations included in the average when computing the average shapes shown in Fig.~\ref{fig:7}. The shapes for the fixed strain rate $\dot{\epsilon}_a=10^{-5}$ and varying spring stiffnesses $k$, shown in Fig.~\ref{fig:7}(a) along with fits of the form of Eq.~\eqref{eq:2} (lines), exhibit a shift in their asymmetry from right to left with decreasing spring stiffness, with $a$ increasing from a slightly negative value for $k=3$ to a large positive value for small $k$. As in case of $k=10$ a pure power law regime cannot be defined, we could not determine a suitable averaging range for that case, and hence the $k=10$ shape is not shown in Fig.~\ref{fig:7}(a). The scaling exponent $\gamma$ obtained from the fits of Eq.~\eqref{eq:2} to the data shows smaller variation with the spring stiffness, however the values $\gamma\approx1.25-1.35$ do not match very well with the $\gamma_\sigma$-exponents obtained for the scaling of the average stress drop magnitudes $\langle \Delta \sigma (T)\rangle$ with $T$ [Fig.~\ref{fig:6}(a)]; this could be due to the rather strong asymmetry distorting the fitted value of $\gamma$, given that Eq.~\eqref{eq:2} assumes the asymmetry to be small. Fig.~\ref{fig:7}(b) shows the average avalanche shapes for the fixed spring stiffness $k=0.1$ and varying the strain rate $\dot{\epsilon}_a$ together with their fits to the model given by Eq.~\eqref{eq:2}. The resulting exponents $\gamma\approx1.25-1.5$ (with $\gamma\approx 1.25$ found for the lowest $\dot{\epsilon}_a$) are again somewhat higher than the scaling exponents of Fig.~\ref{fig:6}(b), similarly to the varying spring stiffness case. All strain rate values result in a leftward asymmetric shape function (positive $a$). Higher strain rates have flatter peaks and temporal asymmetry parameters closer to zero. We note that similar asymmetric avalanche shapes have been observed before in 2D DDD simulations using a different driving protocol~\cite{laurson20061}.

\section{\label{sec:4} Discussion and conclusions}

To conclude, we have presented an extensive study of dislocation avalanches from strain-controlled loading in a simple 2D DDD model. First of all, we found rate dependent deformation such that the average stress at a given strain is higher for a higher imposed strain rate. For strain-controlled loading, the broadly distributed dislocation avalanches are visible as fluctuations around the average stress-strain curves in the form of a sequence of stress drops during which strain is accumulated in bursts. Distributions of both the monotonic stress drops and strain bursts as well as their durations are well-described by power laws with cutoffs.

In general, our results suggest that these distributions might be non-universal such that both the measured exponents as well as the cutoff scales exhibit dependence on $\dot{\epsilon}_a$ and $k$. Interestingly, in the limit of low driving spring stiffness and low imposed strain rate, we found consistently larger exponents for distributions of both stress drops and strain bursts as compared to the exponent of the strain burst distribution found before in simulations employing quasistatic stress-controlled loading~\cite{ispanovity2014avalanches}, although a crossover between the two loading modes was initially expected. We note that this finding of apparently loading mode dependent statistics of dislocation avalanches is in contrast to results from a recent analysis of a mean field model~\cite{PhysRevE.91.042403}, and, more generally, to the scaling picture of depinning phase transitions. In the latter case the avalanche size exponent would be independent of the driving protocol, and only the avalanche size distribution cutoff is affected by the driving parameters. Specifically, for constant velocity driving of elastic interfaces in random media close to the depinning transition, the driving spring stiffness controlling the rate at which the driving force decreases during avalanches is found to tune the cutoff avalanche size while the avalanche size exponent remains unchanged~\cite{laurson2010avalanches}. Here, we have studied a dislocation system which is known to exhibit "glassy features"~\cite{lehtinen2016glassy}, such as "extended criticality" under quasistatic stress-controlled loading~\cite{ispanovity2014avalanches,ovaska2015quenched,salmenjoki2020plastic}, instead of the typical signatures of depinning phase transitions.Here these features are manifested as a larger avalanche size exponent in the limit of small $k$ and $\dot{\epsilon}_a$ than the one found when employing the quasistatic stress-controlled loading where applied stress is kept constant during the avalanches. Our study of quasistatic stress-controlled loading with strain rate thresholding shows, however, that the power law exponent depends on the definition of the avalanche (given by a strain rate threshold in this case), being able to reach values up to $\tau_{\epsilon,\mathrm{INT}} \approx 1.8$ for high strain rate thresholds, comparable to the results of the strain-controlled simulations. This investigation of the avalanche threshold's definition reveals the essential difference between the two loading modes. The governing equation of the strain rate loading sets the imposed strain rate (coinciding by definition with the mean of the strain rate signal in the long-time limit) as the avalanche threshold, while stress-controlled loading is driven by a threshold close to the minimum strain rate. Note that a previous 3D DDD study found $\tau_{\epsilon,\mathrm{INT}} \approx 1.55$ for strain-controlled simulations~\cite{kapetanou2015statistical}, while 3D DDD simulations with quasistatic stress-controlled loading (thresholding the collective velocity signal to define avalanches) found $\tau_{\epsilon,\mathrm{INT}} \approx 1.3$~\cite{lehtinen2016glassy}, similarly to our observation here in the 2D case where the same strain rate threshold was used to define both the driving and the avalanches.

The driving parameter dependent nature of the strain bursts in strain-controlled loading is further highlighted by the evolution of the average strain burst shapes with $\dot{\epsilon}_a$ and $k$. When choosing the averaging range from the power law regime of $P_{\mathrm{INT}}(T)$, the parameters quantifying the average burst shapes, $\gamma$ and $a$ in Eq.~\eqref{eq:2}, are found to depend on $\dot{\epsilon}_a$ and $k$. This is again in contrast to observations in systems where the avalanche dynamics stems from an underlying depinning transition where the average avalanche shapes are "universal" within the power law scaling regime, and distortions of this universal shape are found only for avalanches belonging to the avalanche duration distribution cutoffs~\cite{laurson2013avalancheshape}.

It would be interesting to check to what extent these results are generalizable to 3D DDD simulations~\cite{lehtinen2016glassy}. Moreover, studying avalanches due to strain-controlled loading in DDD simulations with quenched pinning interfering with dislocation motion would also be of interest, given that there the dislocation avalanches have been found to exhibit depinning-like characteristics~\cite{ovaska2015quenched,salmenjoki2020plastic}. The stiffness of the specimen-machine system is not a controllable parameter in typical experiments, but it would be interesting to carefully analyze possible strain rate dependence of avalanche statistics in experiments along the lines proposed in this paper.

\begin{acknowledgments}
The authors acknowledge the support of the Academy of Finland via the Academy Project COPLAST (project no. 322405). 
\end{acknowledgments}

\bibliographystyle{apsrev4-2}
\bibliography{apssamp}

\end{document}